# RESAMPLING APPROACH TO THE ESTIMATION OF THE AIRCRAFT CIRCULATION PLAN RELIABILITY

Maxim FIOSHIN [1]

**Summary.** The paper illustrates an application of the Resampling approach [2] for the estimation of the aircraft circulation plan reliability. Resampling is an intensive computer statistical method, which can be used effectively in the case of small samples. Algorithm of the Resampling method for the given task is illustrated and variance of obtained estimators is calculated, which is the measure of the method effectiveness.

## 1. INTRODUCTION

Reliability and safety of transfers requires accurate analysis of transport functionality models. Many statistical problems appear in this area. One of the main problems in classical methods application for transport models analysis is connected with insufficiency of primary data. In this case, Resampling method can be used efficiently [1].

In the present paper, we consider the Resampling method application to the aircraft circulation plan reliability estimation. An aircraft executes flights from the given airport and returns back. During the given period it should execute a given number of flights. We have statistics about its service times and delays. Our task is to estimate the probability that the aircraft will execute all flights without delays and an aircraft from reserve will not be necessary.

As we have small statistics about delays of the concrete flights in the given period of the year and small statistics about service times in concrete conditions, the Resampling method can be efficiently used for the given task.

In the second section of the paper we will formulate a problem. The third section will illustrate how to apply the Resampling approach for the given problem. The fourth section will illustrate how to calculate variance of obtained estimator, which is the measure of the method efficiency. In the fifth section the numerical example is considered. Conclusions end the paper.

[1] Faculty of Computer Science and Electronics, Transport and Telecommunications Institute, Lomonosov Str. 1, LV1019 Riga, Latvia
 mf@tsi.lv

## 2. PROBLEM FORMULATION

Let us have an aircraft, which executes *k+1* flights from the given airport during the given period of time. Starting at the beginning of the period, it executes the first flight and returns back, then executes the second flight and returns back etc. Times of the aircraft departures are known, and departures should not be delayed. It is called an aircraft circulation plan.

When the aircraft returns from the flight, the technical service is required. It takes some time, and it should be a time for the technical service between the aircraft arrival and departure to the next flight.

Delays of arrivals are possible. If the time between arrival and departure is enough for the delay compensation and necessary service, then the aircraft departs to the next flight according to the timetable. Otherwise an aircraft from reserve should be used to avoid delay of the departure.

Our task is to estimate the probability, that the aircraft will execute all flights without delay. It should be estimated on the base of available statistics about separate flights delays and service times after the given flights.

Let us describe our model more in more formal way. Let $X_1, X_2, ..., X_k$ be mutually independent random variables, which describe delays from corresponding flights. Let us denote $F_1(x), F_2(x),...,F_k(x)$ the distribution functions of these variables.

Let $Y_1, Y_2, ..., Y_k$ be mutually independent random variables, which describe the times of the technical service after corresponding flights. Let us denote $G_1(x), G_2(x),...,G_k(x)$ the distribution functions of these variables.

There is a timetable of flights, or a circulation plan. We will define it by the vector of intervals $T=(t_1, t_2,...,t_k)$ between flights. The time $t_i$ denotes the interval between arrival from the i-th flight according to the timetable and departure to the next, i+1-th flight.

Let $A_T$ be event "the circulation plan T is executed without delays in departures". Let us define $R(T)$ the probability of this event:

$$R(\mathbf{T}) = P\{A_\mathbf{T}\} = P(X_1 + Y_1 \leq t_1, X_2 + Y_2 \leq t_2, ..., X_k + Y_k \leq t_k). \quad (1)$$

Let us define $A_i$ the event "$X_i + Y_i \leq t_i$", which means: "after *i*-th flight the aircraft can depart to the *i+1*-th flight without delay". Let $R_i(t)$ be the probability of this event:

$$R_i(t) = P\{X_i + Y_i \leq t_i\}. \quad (2)$$

As the events $X_1 + Y_1 \leq t_1, X_2 + Y_2 \leq t_2, ..., X_k + Y_k \leq t_k$ are independent, formula (1) can be written in the following way:

$$R(T) = \prod_{i=1}^{k} R_i(t_i). \quad (3)$$

We can calculate the reliability function $R_i(t)$ in the following way:

$$R_i(t) = \int_0^\infty F_i(t-x)dG_i(x). \quad (4)$$

The circulation plan is illustrated on the Figure 1.

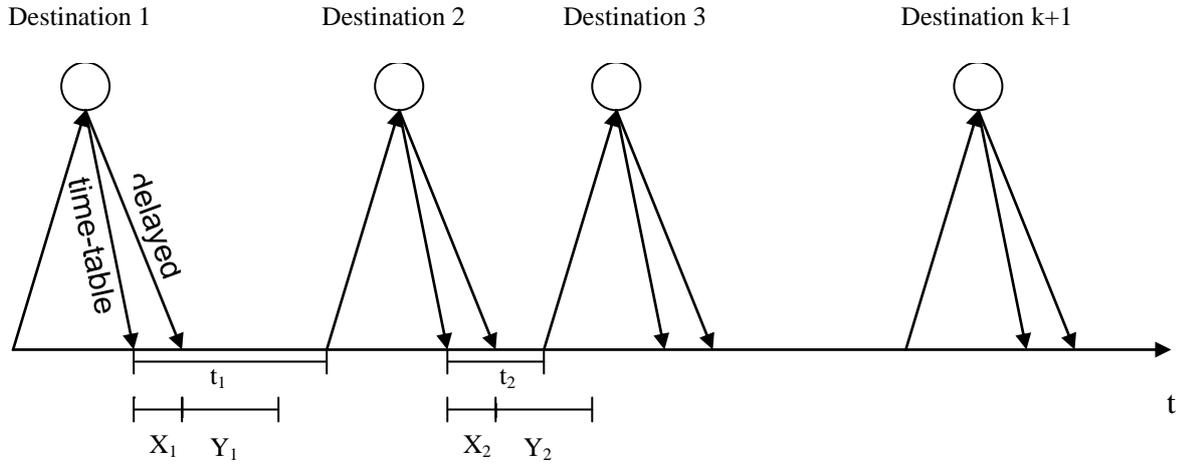

Fig. 1. Circulation plan

Now let us define a function $\phi_T(X,Y)$, where $X$ and $Y$ are vectors of the size $k$: $X = (x_1, x_2, ..., x_k)$, $Y=(y_1, y_2, ..., y_k)$:

$$\phi_T(X,Y) = \begin{cases} 1 & \text{if } x_1 + y_1 \leq t_1, x_2 + y_2 \leq t_2, ..., x_k + y_k \leq t_k, \\ 0 & \text{else.} \end{cases} \quad (5)$$

Clear that the reliability function $R(T)$ is an expectation of the function $\phi_T(X,Y)$, which arguments are vectors of random variables $X=(X_1, X_2, ..., X_k)$ and $Y=(Y_1, Y_2, ..., Y_k)$, and an expectation is taken in respect to vectors $\mathbf{X}$ and $\mathbf{Y}$.

$$R(T) = E_{X,Y}\, \phi_T(X,Y). \quad (6)$$

The distribution functions $F_i(x)$ and $G_i(x)$ of the random variables $X_i$ and $Y_i$ are unknown, but only sample populations are available. Let $H_i^X = (X_{i1}, X_{i2}, ..., X_{in_i^X})$ be the sample population for variable $X_i$, $H_i^Y = (Y_{i1}, Y_{i2}, ..., Y_{in_i^Y})$ be the sample population for variable $Y_i$. Our goal is to calculate the reliability function $R(T)$ using the information available in the samples $H_i^X$ and $H_i^Y$.

$$\theta_T = R(T). \quad (7)$$

## 3. RESAMPLING ESTIMATORS

Resampling approach supposes a number of realizations of the function $\phi_T(X,Y)$. During one realization we extract one value from each sample $H_i^X$ and one value from each sample $H_i^Y$. Let us denote $j_{X,i}(l)$ an index of the element extracted from the sample $H_i^X$ on the *l*-th realization, $j_{Y,i}(l)$ an index of the element extracted from the sample $H_i^Y$ on the *l*-th realization. Let us construct the corresponding vectors $j_X(l) = (j_{X,1}(l), j_{X,2}(l), ..., j_{X,k}(l))$ and $j_Y(l) = (j_{Y,1}(l), j_{Y,2}(l), ..., j_{Y,k}(l))$.

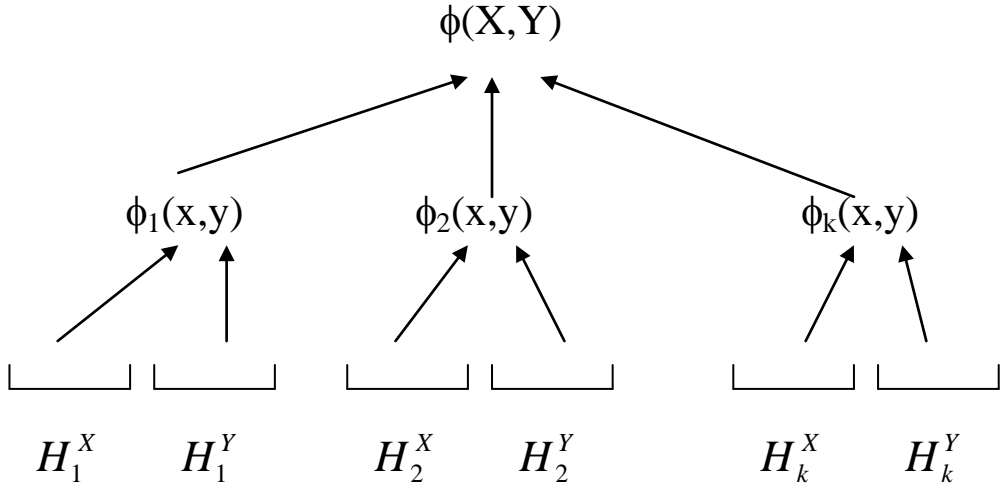

Fig. 2. Resampling procedure

Let us denote $X(l)$ the vector of values extracted on the *l*-th step, $X(l) = (X_{j_{X,1}(l)}, X_{j_{X,2}(l)}, ... X_{j_{X,k}(l)})$ and $Y(l)$ the vector of values extracted on the *l*-th step $Y(l) = (Y_{j_{X,1}(l)}, Y_{j_{X,2}(l)}, ... Y_{j_{X,k}(l)})$.

We repeat this procedure obtaining *r* such realizations. The Resampling estimator $\theta_T^*$ of the parameter $\theta_T$ is calculated as an average on all *r* realizations:

$$\theta_T^* = \frac{1}{r}\sum_{l=1}^{r} \phi(X(l),Y(l)). \qquad (8)$$

As the conservative extraction plan takes place [3], the estimator (8) is unbiased:

$$E_{X,Y}\, \theta_T^* = \theta_T. \qquad (9)$$

So, as the Resampling estimator $\theta_T^*$ is unbiased, we use its variance $D\,\theta_T^*$ as the method's efficiency criteria. We will show how to calculate it in the next section.

## 4. VARIANCE CALCULATION

Let us introduce the following notations: $\mu_\upsilon$ is $\upsilon$-th order moment of the function $\phi_T$ Resampling realization, $\mu_{11}$ is a mixed moment of different function $\phi$ realizations:

$$\mu_\upsilon = E\ \phi_T(X(l),Y(l)),\ \upsilon=1,2,..., \quad (10)$$
$$\mu_{11} = E\ \phi_T(X(l),Y(l))*\phi(X(l'),Y(l')),\ l\neq l'.$$

Using standard formula for variance, we have the following expression for $D\ \theta_T^*$:

$$D\ \theta_T^* = \frac{\mu_2}{r} + \frac{r-1}{r}\mu_{11} - \mu^2. \quad (11)$$

Note that in (11) values $\mu_2$ and $\mu$ are characteristics of the function $\phi_T$ and distributions of its arguments. Only value $\mu_{11}$ depends on the way of elements extraction. Next we will show how to calculate it.

Values $\mu_2$ and $\mu$ can be calculated simply, they are equal to $\theta_T$:

$$\mu_2 = \mu = \theta_T. \quad (12)$$

In order to calculate the mixed moment $\mu_{11}$, we will use the $\omega$-pair notation [2]. We say that two vectors $j_X(l)$ and $j_X(l')$ produce the $\omega$-pair, $\omega \subset \{1,2,...,k\}$ if $j_{X,i}(l) = j_{X,i}(l')$ if and only if $i \in \omega$. The $\omega$-pair notation is illustrated on the Figure 3.

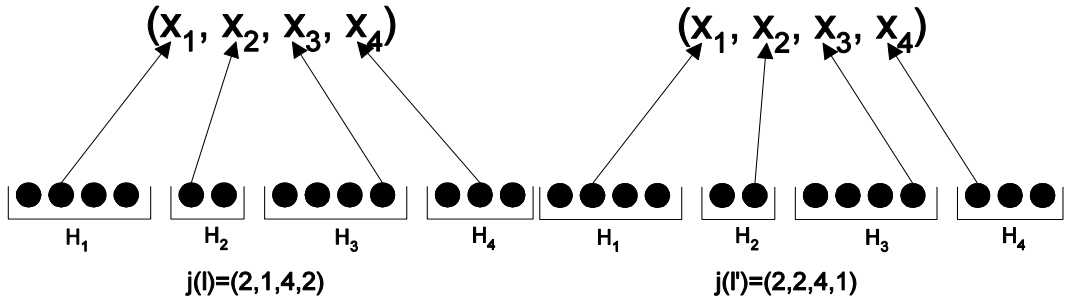

Fig. 3. {1,3}-pair

Analogously, we say that two vectors $j_Y(l)$ and $j_Y(l')$ produce the $\omega$-pair, $\omega \subset \{1,2,...,k\}$ if $j_{Y,i}(l) = j_{Y,i}(l')$ if and only if $i \in \omega$.

Now let us introduce the $\omega_1\omega_2$-pair. We say that two pairs of vectors $(j_X(l), j_Y(l))$ and $(j_X(l'), j_Y(l'))$ produce the $\omega_1\omega_2$-pair, $\omega_1\omega_2 = \omega_1 \times \omega_2 \subset \{1,2,...,k\}^2$, if vectors $j_X(l)$ and $j_X(l')$ produce the $\omega_1$-pair and vectors $j_Y(l)$ and $j_Y(l')$ produce the $\omega_2$-pair.

Let us define $\mu_{11}(\omega_1\omega_2)$ the conditional mixed moment $\mu_{11}$ by the condition that $\omega_1\omega_2$-pair takes place. $\mu_{11}(\omega_1\omega_2)$ can be calculated by the following way:

$$\mu_{11}(\omega_1\omega_2) = \prod_{i=1}^{k} h_i(\omega_1\omega_2), \qquad (13)$$

where the function $h_i(\omega_1\omega_2)$ is defined as follows:

$$h_i(\omega_1\omega_2) = \begin{cases} \int F_i(t-x)dG_i(x), & \text{if } i \in \omega_1 \text{ and } i \in \omega_2, \\ \left(\int F_i(t-x)dG_i(x)\right)^2, & \text{if } i \notin \omega_1 \text{ and } i \notin \omega_2, \\ \int F_i(t-x)^2 dG_i(x), & \text{if } i \notin \omega_1 \text{ and } i \in \omega_2, \\ \int G_i(t-x)^2 dF_i(x), & \text{if } i \in \omega_1 \text{ and } i \notin \omega_2. \end{cases} \qquad (14)$$

Let us define $P(\omega)$ the probability of the $\omega$-pair. It can be calculated as follows:

$$P(\omega) = \prod_{i \in \omega} \frac{1}{n_i} \prod_{i \notin \omega} (1 - \frac{1}{n_i}). \qquad (15)$$

where $n_i = n_i^X$ or $n_i^Y$ depending on the variable which we are working with.

Clear that the probability of $\omega_1\omega_2$-pair is calculated as

$$P(\omega_1\omega_2) = P(\omega_1) P(\omega_2). \qquad (16)$$

Now the mixed moment $\mu_{11}$ can be calculated as following:

$$\mu_{11} = \sum_{\omega_1,\omega_2} \mu_{11}(\omega_1\omega_2) P(\omega_1\omega_2). \qquad (17)$$

## 5. NUMERICAL EXAMPLE

Let us consider a numerical example. Let us suppose that delays $X_i$ and service times $Y_i$ has an exponential distribution with parameters $\lambda_i$ and $\mu_i$ correspondingly. In this case, the formula (14) can be presented in the following way:

$$h_i(\omega_1\omega_2) = \begin{cases} 1 - \dfrac{\lambda_i e^{-\mu_i t} - \mu_i e^{-\lambda_i t}}{\lambda_i - \mu_i}, & \text{if } i \in \omega_1 \text{ and } i \in \omega_2, \\ \left(1 - \dfrac{\lambda_i e^{-\mu_i t} - \mu_i e^{-\lambda_i t}}{\lambda_i - \mu_i}\right)^2, & \text{if } i \notin \omega_1 \text{ and } i \notin \omega_2, \\ 1 - \dfrac{2\lambda_i^2 e^{-\mu_i t} - 2(2\lambda_i - \mu_i)\mu_i e^{-\lambda_i t} + \mu_i(\lambda_i - \mu_i)e^{-2\lambda_i t}}{(\lambda_i - \mu_i)(2\lambda_i - \mu_i)\mu_i}, & \text{if } i \notin \omega_1 \text{ and } i \in \omega_2, \\ 1 - \dfrac{2\mu_i^2 e^{-\lambda_i t} - 2(2\mu_i - \lambda_i)\lambda_i e^{-\mu_i t} + \lambda_i(\mu_i - \lambda_i)e^{-2\mu_i t}}{(\mu_i - \lambda_i)(2\mu_i - \lambda_i)\lambda_i}, & \text{if } i \in \omega_1 \text{ and } i \notin \omega_2. \end{cases} \qquad (18)$$

Let us take the following values of parameters: flights count $k=5$, parameters $\lambda_i = 0.05$, $\mu_i = 0.02$, sample sizes $n^X_i = n^Y_i = 20$, resamples count $r=50$. We take all times $t_i$ equal to $t$. We will analyze the variance (11) dependence on the time $t$. The results are presented in the Table 1.

Table 1

Variance $D\,\theta_T^*$ of the Resampling estimator

| Nr. | Time, t | Variance, $D\,\theta_T^*$ |
|---|---|---|
| 1 | 20 | $6{,}9\ 10^{-7}$ |
| 2 | 60 | 0,0011 |
| 3 | 100 | 0,0088 |
| 4 | 140 | 0,0124 |
| 5 | 180 | 0,009 |
| 6 | 220 | 0,005 |
| 7 | 260 | 0,0025 |
| 8 | 300 | 0,0011 |

The results are also presented on the Figure 4

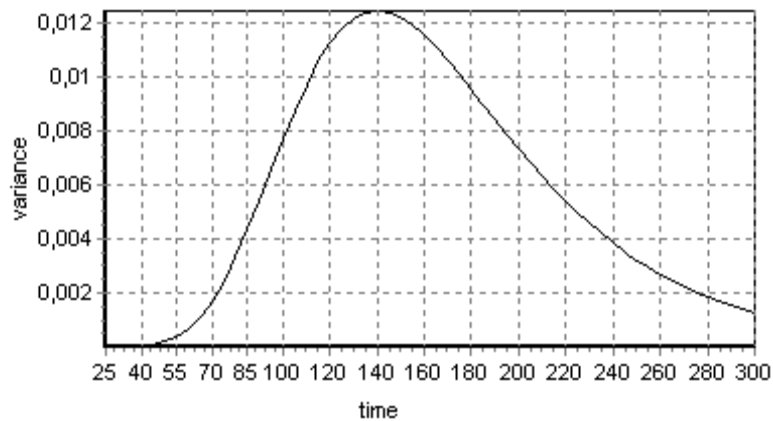

Fig. 4. Variance $D\,\theta_T^*$ dependence on time t

Clear that the behaviour of the variance $D\,\theta_T^*$ depending on the time t is easy to predict: at the beginning we can almost precisely say that it will not be enough time for service and the variance is small; then the variance grows because the probability that it will be enough time also grows; and when the time is big we can almost precisely say that it will be enough time for the service and the variance is small again.

## 6. CONCLUSONS

The proposed approach can be used efficiently in the illustrated case. Obtained formulas allow calculating variance of the estimator. It is shown that the Resampling method can be good alternative to traditional methods.